\documentclass[onecolumn,amsmath,amssymb,12pt,superscriptaddress,nofootinbib,floatfix]{revtex4}
\pdfoutput=1

\usepackage[latin1]{inputenc}
\usepackage[english]{babel}
\usepackage{amssymb}
\usepackage{amsmath}
\usepackage{amsthm}
\usepackage[]{graphicx}
\usepackage[]{subfigure}
\usepackage{tensor}
\usepackage{color}
\usepackage{cancel}
\usepackage{setspace}
\usepackage{fancyhdr}
\usepackage[bookmarks,linktocpage, colorlinks=true, plainpages = false, citecolor = blue,  linkcolor=blue, urlcolor = blue, filecolor = blue]{hyperref}

\def\cB{{\cal B}}

\def\1p{{(1p)}}

\def\p0{\phi_0}

\def\be{\begin{equation}}
\def\ee{\end{equation}}
\def\beq{\begin{eqnarray}}
\def\eeq{\end{eqnarray}}

\newcommand{\pkt}{\; .}
\newcommand{\kma}{\; ,}

\def\e{{\rm e}}

\begin{document}

\title{Aspects of the negative mode problem \\ in quantum tunneling with gravity }
\author{Sebastian F. Bramberger}
\email{sebastian.bramberger@aei.mpg.de}
\affiliation{Max Planck Institute for Gravitational Physics \\ (Albert Einstein Institute), 14476 Potsdam-Golm, Germany}
\author{Mariam Chitishvili}
\email{mariamchitishvili@gmail.com}
\affiliation{I.Javakhishvili Tbilisi State University, GE-0179 Tbilisi, Georgia}
\author{George Lavrelashvili}
\email{george.lavrelashvili@tsu.ge}
\affiliation{Department of Theoretical Physics, A.Razmadze Mathematical Institute \\
	I.Javakhishvili Tbilisi State University, GE-0177 Tbilisi, Georgia}

\begin{abstract}
\vspace{1cm}
\noindent
Some solutions describing vacuum decay exhibit a catastrophic instability. This, so-called negative mode problem in quantum tunneling with gravity, was discovered 34 years ago \cite{Lavrelashvili:1985vn}
and in spite of the fact that in these years
many different groups worked on this topic \cite{Tanaka:1992zw, Khvedelidze:2000cp, Lavrelashvili:1999sr, Gratton:2000fj, Hackworth:2004xb, Lavrelashvili:2006cv, Dunne:2006bt, Yang:2012cu, Battarra:2012vu, Lee:2014uza, Koehn:2015hga, Gregory:2018bdt}, it has still not been resolved. Here, we briefly summarize the current status of the problem and investigate properties of the bounces, numerically and analytically for physically interesting potentials.
In the framework of the Hamiltonian approach \cite{Khvedelidze:2000cp, Gratton:2000fj} we
show that for generic polynomial potentials the negative mode problem could arise at energies much lower than the Planck mass, indicating that the negative mode problem is not related to physics at the Planck scale. At the same time we find that for a Higgs like potential, as it appears in the standard model, the problem does not appear at realistic values of the potential's parameters but only at the Planck scale.
\end{abstract}

\maketitle

\newpage
	
\section{Introduction}

Calculating the decay rate of metastable vacua while taking gravitational effects into account, has risen in importance upon the discovery that we might be living in a false vacuum. Using the Euclidean approach \cite{Coleman:1977py,Callan:1977pt,Coleman:1980aw} for calculating the decay rate of metastable vacua to their true value, $\gamma$, the Arrhenius formula is given by
\be \label{eq:rate}
\gamma = {\cal A} \e^{-{\cal B}} \kma
\ee
with
\be
\cB =S^{(cl)}(\varphi^b)-S^{(cl)}(\varphi^f)\kma
\ee
where the first term on the r.h.s. is the classical Euclidean action calculated along the bounce solution
and the second term is the value of action evaluated at the false vacuum.

\begin{figure}[ht!]
	\centering
	\includegraphics[width=0.5\textwidth]{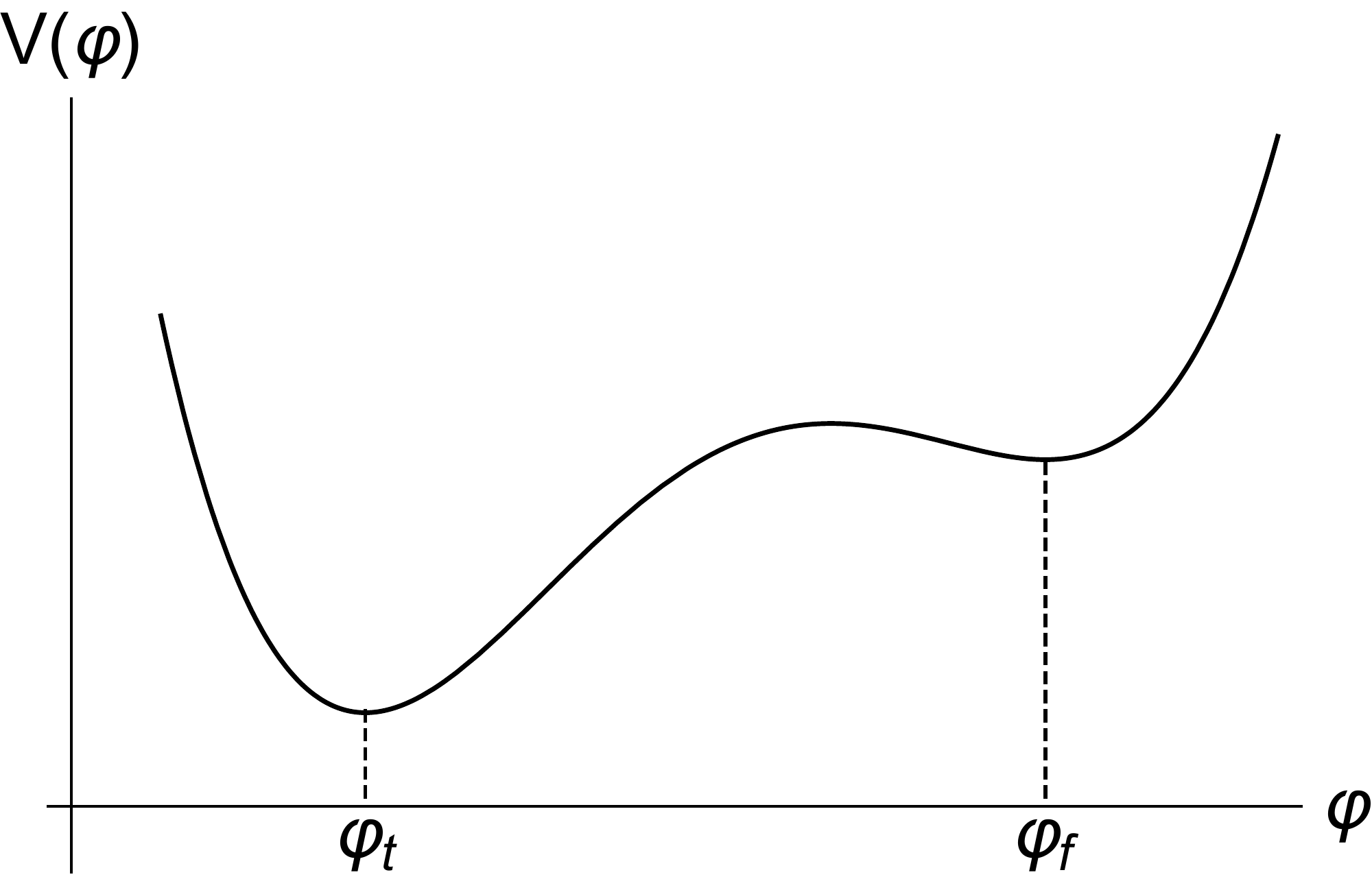}
	\caption{A typical potential in which false vacuum decay can occur. The bounce solution interpolates between the false vacuum $\varphi_f$ and true vacuum $\varphi_t$.
	} \label{fig-false-vac-decay}
\end{figure}

The bounce solution is the lowest action $O(4)$ symmetric solution to the Euclidean equations of motion that interpolates between false and true vacua (see Fig. \ref{fig-false-vac-decay}). Expanding around the bounce solution, gives the pre-exponential factor ${\cal A}$ as a Gaussian integral over the linear perturbations. Proper bounces should have exactly one eigenfunction with a negative eigenvalue in the spectrum of linear perturbations, in order to make the decay picture coherent \cite{Coleman:1987rm}. While this is always the case in flat space-time, generalizing to curved space-time results in some bounces getting infinitely many negative modes indicating a problem. Note that when gravity is involved, in addition to the basic bounce solution, there are oscillating instantons and an infinite tower of oscillating bounces
\cite{Hackworth:2004xb,Lee:2009bp,Lee:2011ms},
which, however, have more than one negative modes \cite{Lavrelashvili:2006cv,Battarra:2012vu} making their relation to tunneling questionable.

Using new approximate analytic methods and numerical calculations, we aim to clarify the question of whether the negative mode problem is inherently related to Planck-scale physics and highlight differences between the Hamiltonian and Lagrangian approaches to the problem. The paper is organized as follows: In the next section we briefly summarize the negative mode problem. In Sec. \ref{poly} we discuss generic quartic polynomial potentials, while in Sec. \ref{Higgs}
we consider a realistic, Higgs-like potential. Finally, the last section contains a summary and concluding remarks.

\section{A short summary of the negative mode problem}

Let's consider the theory of a single scalar field minimally coupled to gravity,
which is defined by the following Euclidean action
\be
S_E=\int {d^4x\sqrt{g} \; \Bigl(-\frac{1}{2\kappa}R + \frac{1}{2}\nabla_\mu\varphi \nabla^\mu\varphi+ V(\varphi) \Bigr)} \kma
\ee
where $\kappa=8\pi G_{N}$ is the reduced Newton's gravitational constant. The most general $O(4)$ invariant metric is parametrised as
\be\label{metric}
ds^2=N^2(\eta)d\eta^2+\rho^2(\eta)d\Omega_3^2 \kma
\ee
where $N(\eta)$ is the lapse function, $\rho(\eta)$ is the scale factor
and $d\Omega_3^2$ is metric of the unit three-sphere.
In proper-time gauge, $N=1,$ the corresponding field equations are
\be \label{eq:phi}
\ddot{\varphi}+3\frac{\dot{\rho}}{\rho}\dot{\varphi}=\frac{\partial V}{\partial\varphi} \kma
\ee
\be \label{eq:rho}
\ddot{\rho}=-\frac{\kappa \rho}{3} \left(\dot{\varphi}^2+V(\varphi) \right) \kma
\ee
\be
\dot{\rho}^2= 1+ \frac{\kappa \rho^2}{3} \left(\frac{\dot{\varphi}^2}{2}-V(\varphi) \right) \kma
\ee
where $\dot{} = d/ d\eta$.
The leading exponential factor in the decay rate is determined by the bounce: A solution of these equations
with appropriate boundary conditions. In order to calculate the pre-exponential factor ${\cal A}$
in Eq.~(\ref{eq:rate}) one should consider linear perturbations about the bounce solution.
For this purpose we expand the metric and the scalar field over an $O(4)$ symmetric background as follows:
\be \label{eq:homogen_metric}
ds^2=  (1+2 A(\eta))d\eta^2 + \rho(\eta)^2 (1-2 \Psi(\eta))d\Omega_3^2
\kma \qquad \varphi=\varphi(\eta) + \Phi(\eta) \kma
\ee
where $\rho$ and $\varphi$ are the background field values and
$A, \Psi$ and $\Phi$ are small perturbations.
Note that under the infinitesimal shift ${\eta \to \eta + \alpha}$ the gauge transformations are
\be \label{eq:gauge transformation}
\delta\Psi =- \frac{\dot{\rho}}{\rho} \alpha \; , \qquad
\delta\Phi = \dot{\varphi} \alpha \;, \qquad
\delta A =  \dot{\alpha}  \; \pkt
\ee

In what follows, we will be interested in the lowest (purely $\eta$-dependent, `homogeneous') modes
and consider only scalar metric perturbations. Expanding the total action to second order in perturbations and using the background equations of motion, we find
\be
S= S^{(0)}[\rho,\varphi] + S^{(2)}[A,\Psi,\Phi] \kma
\ee
where $S^{(0)}$ is the action of the background solution and $S^{(2)}[A,\Psi,\Phi]$
is the quadratic action.
An analysis of the equations of motion following from this quadratic action shows \cite{Lavrelashvili:1985vn,Lavrelashvili:1999sr} that there are constraints in this system and only one out of three variables is physical. The unconstrained quadratic action about Coleman - De Luccia bounces was first derived in \cite{Lavrelashvili:1985vn} using the $\Psi = 0$ gauge in the Lagrangian approach. Integrating out $A$ and expressing the quadratic action in terms of the remaining, physical perturbation $\Phi$, one gets
\be
S^{(2)}_{L}= 2 \pi^2 \int \rho^3 d\eta \left[ \frac{\dot{\rho}^2}{2 Q_{L}}  \dot{\Phi}^2 + \frac{1}{2} U_\Phi \Phi^2 \right]
\ee
with the potential being
\be
U_\Phi = \frac{\dot{\rho}^2 V''}{Q_{L}}  + \frac{\kappa \rho^2 \dot{\rho}^2 V'^2}{3 Q_{L}^2}+\frac{\kappa \rho \dot{\rho} \dot{\varphi} V' }{3 Q_{L}^2} \kma
\ee
where $' \equiv d/d\varphi$.
In particular, it was noted that a factor termed $Q$ appears in front of the kinetic term, which in the Lagrangian approach is the following combination of background quantities
\be \label{eq:Qlrt}
Q_{L} = 1-\frac{\kappa \rho^2 V(\varphi)}{3}=\dot{\rho}^2 -\frac{\kappa \rho^2 \dot{\varphi}^2}{6} \pkt
\ee
This factor becomes negative for any bounce solution close to the point $\dot{\rho} =0$. In addition, for some bounces it becomes negative a second time, in a regime where the last term dominates over $\dot{\rho}$. Despite its widespread use, the Lagrangian approach was criticized in \cite{Tanaka:1992zw} because of poor gauge fixing. Indeed, from the gauge transformations Eq.~(\ref{eq:gauge transformation}) it is clear that we cannot freely transform the variable $\Psi$. In particular the transformation breaks down at any point where $\dot{\rho}=0$ making it impossible to impose a nonsingular gauge on $\Psi$. Unfortunately, there are not many alternatives in the Lagrangian approach since it only involves configuration space variables. Later, Lee and Weinberg \cite{Lee:2014uza} promoted $\Phi$ to a gauge invariant variable
\be
\chi = \dot{\rho} \Phi +\rho \dot{\varphi} \Psi \kma
\ee
and obtained a pulsation equation, which exactly coincides with the earlier $\Psi = 0$ gauge fixed
approach (see Appendix in \cite{Koehn:2015hga}).

Therefore, we will use the Hamiltonian approach in this note which is more adequate for constrained dynamical systems. Using a Hamiltonian approach following Dirac the quadratic action has the form \cite{Khvedelidze:2000cp, Koehn:2015hga}
\be
S^{(2)}_{H}= \pi^2 \int d\eta \Phi \left[-\frac{d}{d\eta} \left(\frac{\rho^3(\eta)}{Q_H}\frac{d}{d\eta} \right)
+ \rho^3(\eta) U[\varphi(\eta),\rho(\eta)] \right] \Phi \kma
\ee
where the potential $U$ is expressed in terms of the bounce solution as
\be 
U[\varphi(\eta),\rho(\eta)] \equiv \frac{V''(\varphi)}{Q_H}+\frac{2\kappa{\dot{\varphi}}^2}{Q_H}
+\frac{\kappa}{3 Q_H^2} \Bigl(6{\dot{\rho}}^2{\dot{\varphi}}^2 +\rho^2 V'^2(\varphi)
-5\rho \dot{\rho}\dot{\varphi} V'(\varphi)\Bigr) \pkt
\ee
and again a factor $Q_{H}\equiv Q$ appears in quadratic action and this time it reads
\be \label{eq:Q}
Q = 1-\frac{\kappa \rho^2 \dot{\varphi}^2}{6} \pkt
\ee
Unlike the previous prefactor in Eq. (\ref{eq:Qlrt}), this factor is positive definite for a wide class of bounces where one finds exactly one {\it tunneling} negative mode in the spectrum of the unconstrained action \cite{Khvedelidze:2000cp,Lavrelashvili:1999sr,Gratton:2000fj,Koehn:2015hga}.
When $Q$ becomes negative along the bounce,
the pulsation equation is regular and the tunneling negative mode persists,
but on top of it one gets an infinite tower of negative modes that has support in the negative $Q$ region. Furthermore, negative $Q$ leads to catastrophic particle creation and instability of the
quasiclassical approximation \cite{Lavrelashvili:1985vn}.

\section{Negative mode problem for a polynomial potential}\label{poly}
\subsection{Numerical example of negative Q far from Planck scale}
\begin{figure}[ht!]
	\centering
	\includegraphics[width=0.5\textwidth]{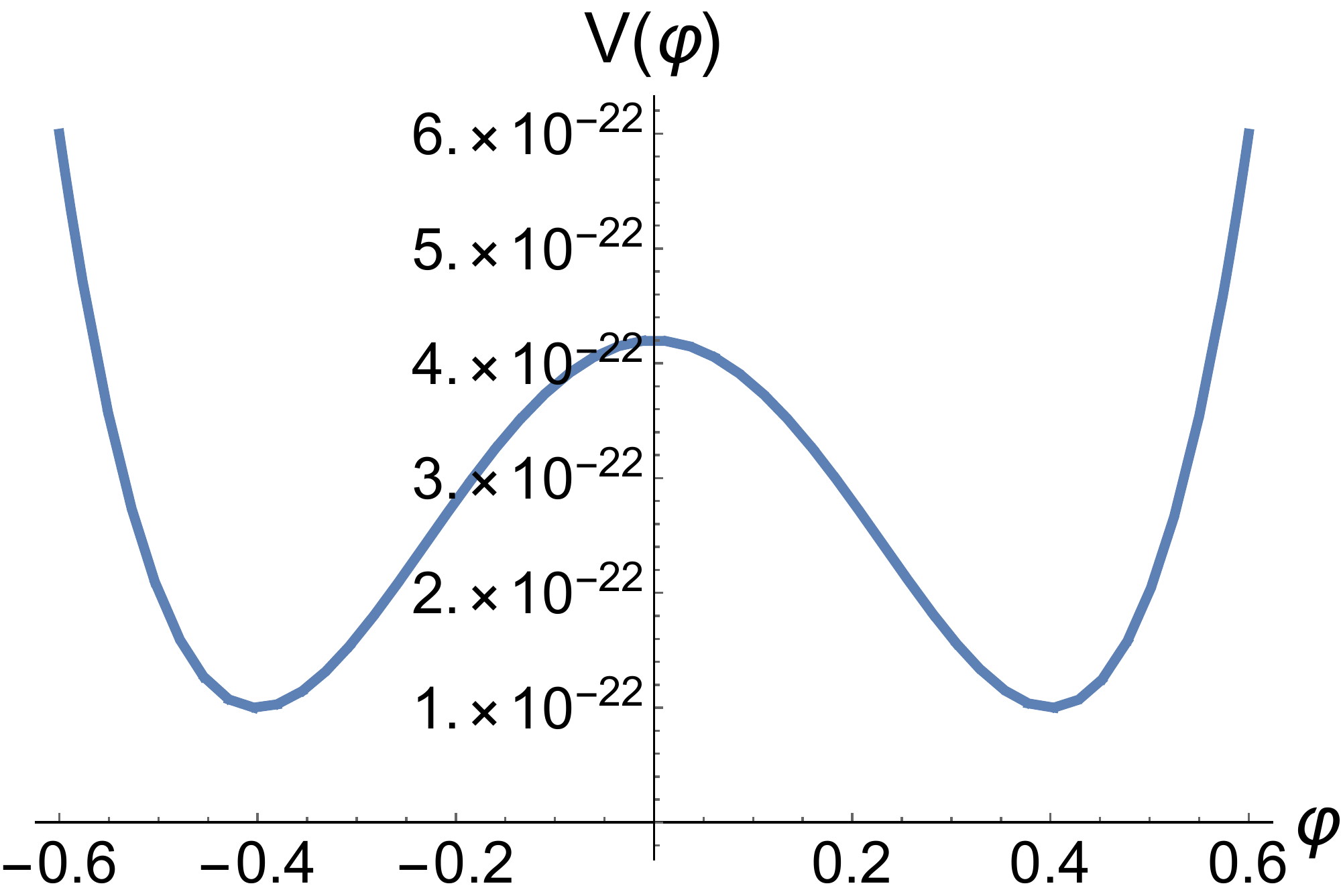}
	\caption{A plot of the potential Eq.(\ref{eq:quartic}) for the parameter values  $V_0 = 10^{-22}$, $\lambda = 10^{-19}$, $\epsilon = 10^{-30}$,
and $\mu = 0.4$. For these parameters we have $V(\varphi_{top})$ five orders of magnitude below the Planck scale. The minima for this potential are almost degenerate, a fact, which is reflected in the small value for $\epsilon$, but there still is a true and a false vacuum.
	} \label{fig-potential}
\end{figure}

\begin{figure}[ht!]
	\centering
	\includegraphics[width=0.40\textwidth]{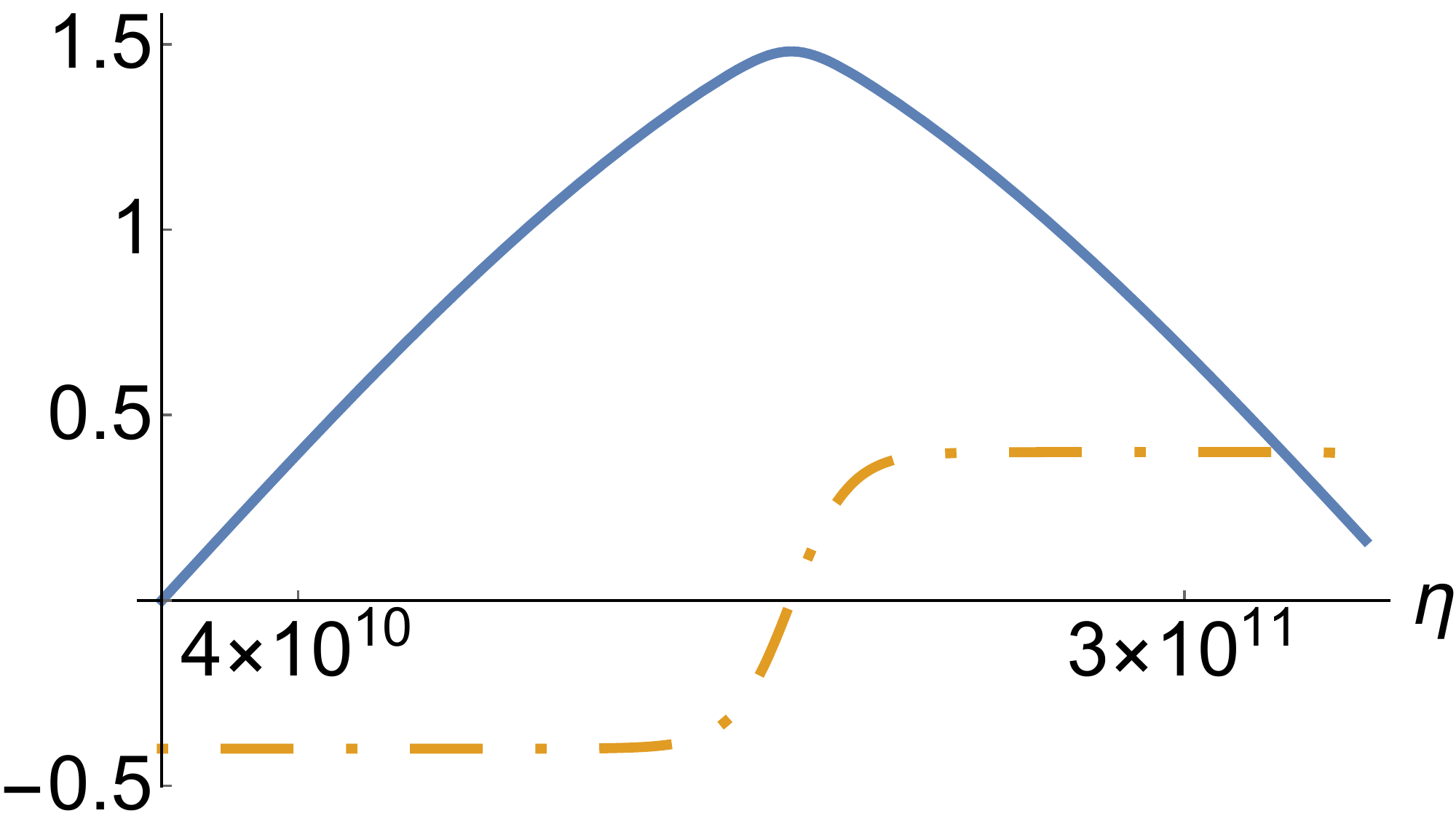} 
	\hspace{0.1\textwidth}
	\includegraphics[width=0.40\textwidth]{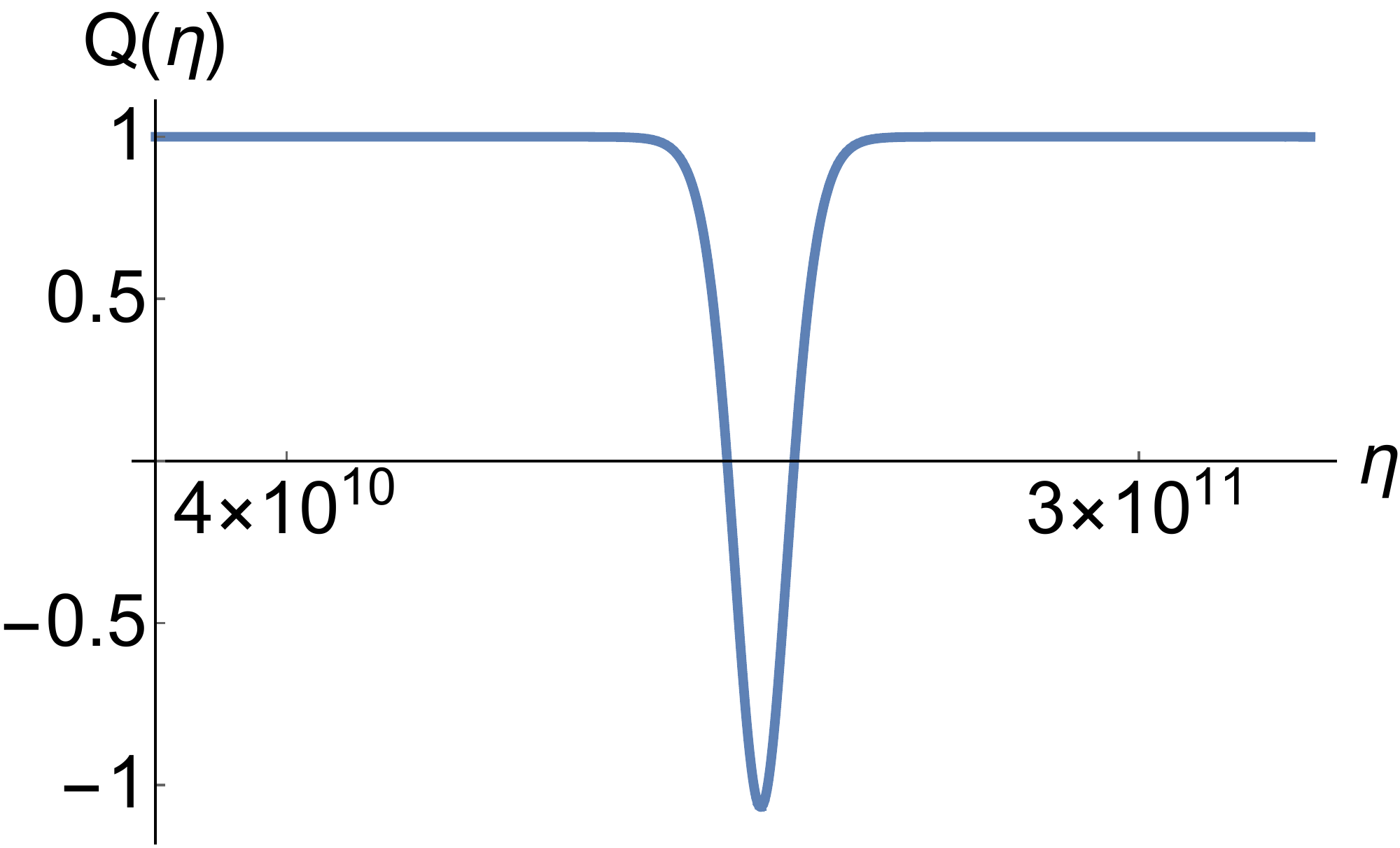}
	\caption{\textit{Left:} The evolution of the scale factor $\rho(\eta)/10^{11}$ in blue and scalar field $\varphi(\eta)$ in orange as a function of Euclidean time $\eta$ which ranges from 0 to approximately $3.6 \times 10^{11}$ in this example. 
		\textit{Right:} The evolution of Q for this instanton clearly demonstrating that it becomes negative along the bounce solution.} \label{fig-CdL-instanton}
\end{figure}

\begin{figure}[ht!]
	\centering
	\includegraphics[width=0.5\textwidth]{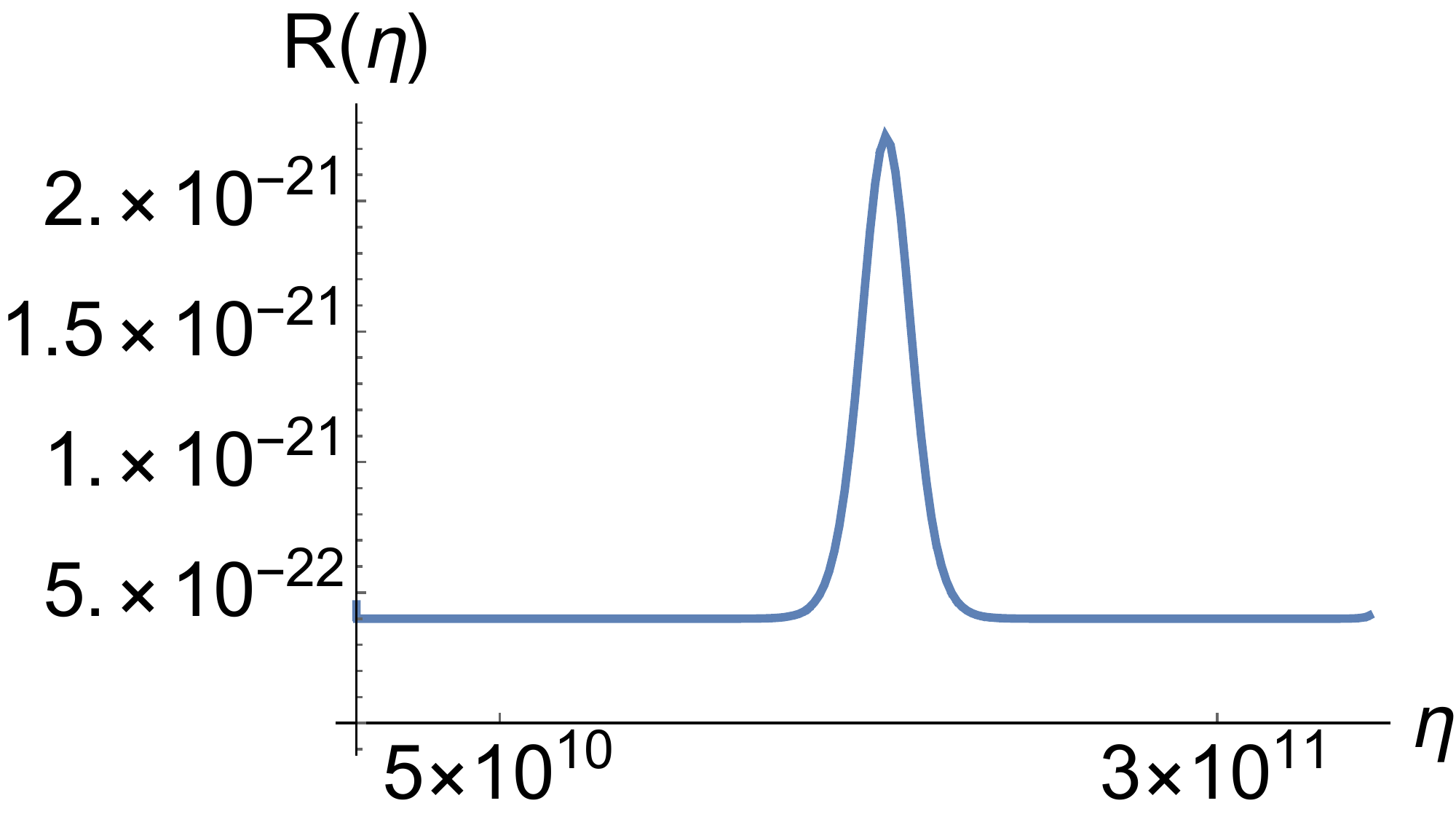}
	\caption{The four dimensional Ricci scalar for the instanton solution in Fig.~\ref{fig-CdL-instanton}}. \label{fig:curvature}
\end{figure}

One might argue that the problematic behaviour of $Q$ only appears close or above the Planck scale where classical General Relativity is no longer valid. 
Here with combined numerical and analytic methods we can show that this is not the case and $Q$ may be negative even far away from the Planck scale. For definiteness we parameterize the quartic potential as
\begin{align}\label{eq:quartic}
V(\varphi) = V_0 + \frac{\lambda}{8}(\varphi^2 - \mu^2)^2 + \frac{\epsilon}{2\mu}(\varphi + \mu)
\end{align}
and plot it in Fig.~\ref{fig-potential}.
The evolution of the scale factor and scalar field for the Coleman - De Luccia bounce solution and the evolution of the corresponding $Q$ factor is shown in Fig.~\ref{fig-CdL-instanton} and we can immediately see that even though the energy scale is significantly below the Planck scale, $Q$ turns negative along the evolution. It might be argued that $Q$ becomes negative because the curvature becomes huge close at the maximal radius of the instanton. However, the four-dimensional Ricci scalar $R$, given by
\begin{align}
R = \frac{6}{\rho(\eta)^2} \left( 1 - \dot{\rho}(\eta)^2 - \rho(\eta) \ddot{\rho}(\eta) \right)
\end{align} is suppressed by a factor of $\frac{1}{\rho^2}$, where the scale factor $\rho$ typically is large in the negative $Q$ regime. Hence, the curvature is expected to be small as well which is demonstrated for the example above in Fig.~\ref{fig:curvature}. In general the intuitive reasoning of $\varphi$ rolling in the inverted potential gives a good guideline for how to find solutions with negative Q at an arbitrary scale. In particular, taking $V(\varphi_{top})$ much bigger than $V(\varphi_{\pm})$ where $\varphi_{\pm}$ are the two deSitter vacua of the potential will give a fast rolling field with a large bubble radius which are the exact conditions for negative $Q$. In the next section we make this argument more precise.

\subsection{Negative Q in the thin wall approximation}

 We are interested in a formula for $Q$ that depends only on the parameters of the potential. Critically we note that the smallest value of $Q$
 (see Eq.~(\ref{eq:Q})) is obtained when $\rho^2 \dot{\varphi}^2$ is maximized which, in the thin wall limit approximately happens when both $\rho$ and $\dot{\varphi}$ are extremized. Thus, starting with $\rho$, the general formula for the bubble size \cite{Parke:1982pm} is
\begin{align} \label{eq:park}
\rho^2 = \frac{\rho_0^2 }{1 + 2 (\rho_0^2/2{\bar \lambda})^2 + (\rho_0/2{\bar \Lambda})^4} \kma
\end{align}
where $\epsilon$ is the separation between the true and false vacuum $\epsilon = V_f - V_t$, $\rho_0$ is the critical bubble size without gravity and
\begin{align} \label{eq:definitions}
{\bar \lambda}^2 = \frac{3}{\kappa(V_f + V_t)} = \frac{3}{\kappa (2V_f - \epsilon)}, \indent
{\bar \Lambda}^2 = \frac{3}{\kappa(V_f - V_t)} \pkt
\end{align}
This provides a generalization of Coleman - De Luccia's earlier result which can be recovered by setting $\bar{\Lambda}^2/\bar{\lambda}^2 = \pm 1$
corresponding to $V_f = 0$ or $V_t = 0$ respectively.
Using definitions Eq.~(\ref{eq:definitions}), expression for bubble size Eq.~(\ref{eq:park}) can be written as follows
\begin{align}
\rho^2 = \frac{\rho_0^2 }{\frac{\kappa \rho_0^2 V_f}{3} + \left(1 - \frac{\kappa \rho_0^2 \epsilon}{12}\right)^2} \pkt
\end{align}
This expression shows that in contrast to flat space-time, where bubble size grows indefinitely when $\epsilon \to 0$,
in dS-dS transition it reaches maximum size and starts to decrease again.
Hence this expression simplifies dramatically by taking a particular value for $\epsilon$, namely
\begin{align}
\epsilon = \frac{12}{\kappa \rho_0^2 } = \frac{3}{4}\kappa \sigma^2 \kma
\end{align}
where $\sigma$ is the bubble tension in the absence of gravity. Due to this choice the bubble size now takes on a particularly simple form
\begin{align}
\rho^2 = \frac{3}{\kappa V_f} \pkt
\end{align}
So far all the calculations were independent of the particular form of the potential. One can go one step further and obtain a concrete value for $\epsilon$ based on the parameters of the potential by choosing
\begin{align} \label{eq:quartic-potential}
V(\phi) = \frac{c^2}{8}(\varphi^2 - \mu^2)^2 + \frac{\epsilon}{2 \mu} \left( \varphi + \mu\right) \kma
\end{align}
where $c^2 > 0, \mu > 0$ and $\epsilon \geq 0$, such that the wall tension $\sigma$ can be solved for analytically, in the thin wall approximation
\begin{align}
\sigma = \int_{\varphi_t}^{\varphi_f} \left[2 \left( V_s(\varphi) - V_s(\varphi_t) \right) \right]^{1/2}  d \varphi = \frac{2}{3}c\mu^3 \kma
\end{align}
where $V_s = \frac{\lambda}{8}(\varphi^2 - \mu^2)^2$ is the symmetric part of the potential and for this potential we have $\varphi_{t,f} = \pm \mu$. This implies that the critical value for $\epsilon$ is
\begin{align} \label{eq:special-eps}
\epsilon = \frac{1}{3}\kappa c^2 \mu^6 \pkt
\end{align}
Returning to the definition of $Q$ and making use of the Friedman equation
\begin{align}
\dot{\rho}^2 = 1 + \frac{\kappa}{3} \rho^2 \left( \frac{1}{2}\dot{\varphi}^2 - V(\varphi) \right)
\end{align}
we obtain
\begin{align}
Q &= 2 - \dot{\rho}^2 - \frac{\kappa}{3} \rho^2 V(\varphi)
\end{align}
and consequently, if we restrict $\epsilon$ to be of the special form of Eq.~(\ref{eq:special-eps}), we have
\begin{align}
Q_c &= 2 - \dot{\rho}^2 - \frac{V(\varphi)}{V_f} \qquad \rightarrow  \qquad Q_c \leq 2 - \frac{V(\varphi)}{V_f} \pkt
\end{align}
Hence if we can find a $\phi$ such that this quantity is negative, we can be sure that $Q$ will be negative somewhere. As a first guess we can take for example $\phi_c = 0$. Numerically we will see that this assumption leaves us very close to the extremal value for $Q_c$. Writing this in terms of the parameter of the potential given in Eq.~(\ref{eq:quartic-potential}), we obtain:
\begin{align}
Q_c & \leq 2 - \frac{V(\varphi)}{V_f} \approx 2 - \frac{V(0)}{V_f} \\
&= 2 - \frac{1}{V_f} \left( \frac{c^2}{8} \mu^4+ \frac{\epsilon}{2}\right) \\
&\approx  \frac{3}{2} -  \frac{c^2}{8} \frac{\mu^4}{\epsilon} \\
&=  \frac{3}{2} \left(1 - \frac{1}{4 \kappa \mu^2} \right)
\end{align}
where in the last approximation we took $\varphi_t \approx \mu$ which implies $V_f \approx \epsilon$ and we have plugged in the critical value for epsilon in the second last line. All this implies that for $  \mu^2 < \frac{1}{4\kappa} $ we expect that Q is negative at some point. This confirms our intuition that for steeper potentials we expect Q to be more negative since the scalar field will roll faster in such a potential. Indeed, this choice of $\epsilon$ illustrates this beautifully since it eliminates the dependence on the height of the potential. Thus we can find transitions that have the problematic negative pre-factor for the kinetic term of the perturbations at $any$ scale.

\subsection{Existence of Coleman - De Luccia solutions}

It is known  \cite{Jensen:1983ac}, \cite{Hackworth:2004xb} that for the existence of Coleman - De Luccia bounce solution 
in a given potential $V(\varphi)$ following condition should be satisfied
\begin{align}
|V''(\varphi_{top})| > 4H^2 (\varphi_{top}) \kma
\end{align}
where $V''(\varphi)= \frac{d^2 V(\varphi)}{d\varphi^2}$ and $H^2 (\varphi) = \frac{\kappa V(\varphi)}{3}$. For the quartic potential defined in Eq.~(\ref{eq:quartic-potential})
we approximate $\varphi_{top} = 0$ and consequently must satisfy
\begin{align}
\frac{c^2\mu^2}{2} > \frac{2}{3} \kappa \left( \frac{c^2 \mu^4}{4} + \epsilon \right)
\end{align}
Choosing $\epsilon = \frac{1}{3}\kappa c^2 \mu^6$, as above, we find that in order for Coleman - De Luccia instantons to exist we must have
\begin{align}
\mu^2 < \frac{3}{8\kappa}(\sqrt{17}-1) \approx \frac{9}{8 \kappa }
\end{align}
Hence for $0 < \mu^2 < \frac{1}{4\kappa}$, Coleman - De Luccia
solutions exist but are pathological as $Q$ is negative for some part of the instanton. For $\frac{1}{4\kappa} < \mu^2 < \frac{9}{8\kappa}$, 
the Coleman - De Luccia instantons exist and are perfectly well behaved while for $\mu^2 > \frac{9}{8\kappa}$ no 
Coleman - De Luccia solutions exist.

\subsection{Comparison with numerics}
\begin{figure}[ht!]
	\centering
	\includegraphics[width=0.45\textwidth]{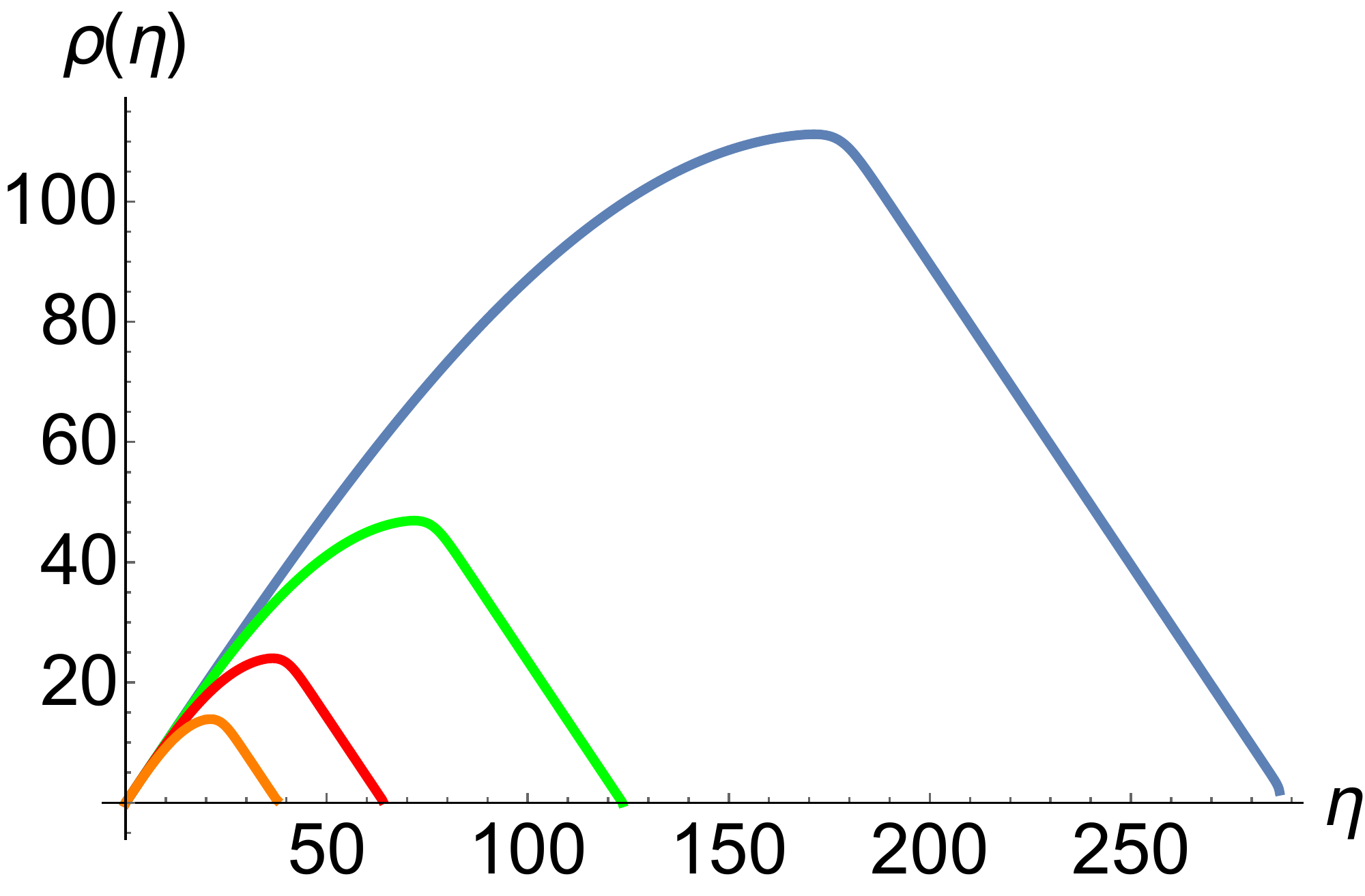}
	\includegraphics[width=0.45\textwidth]{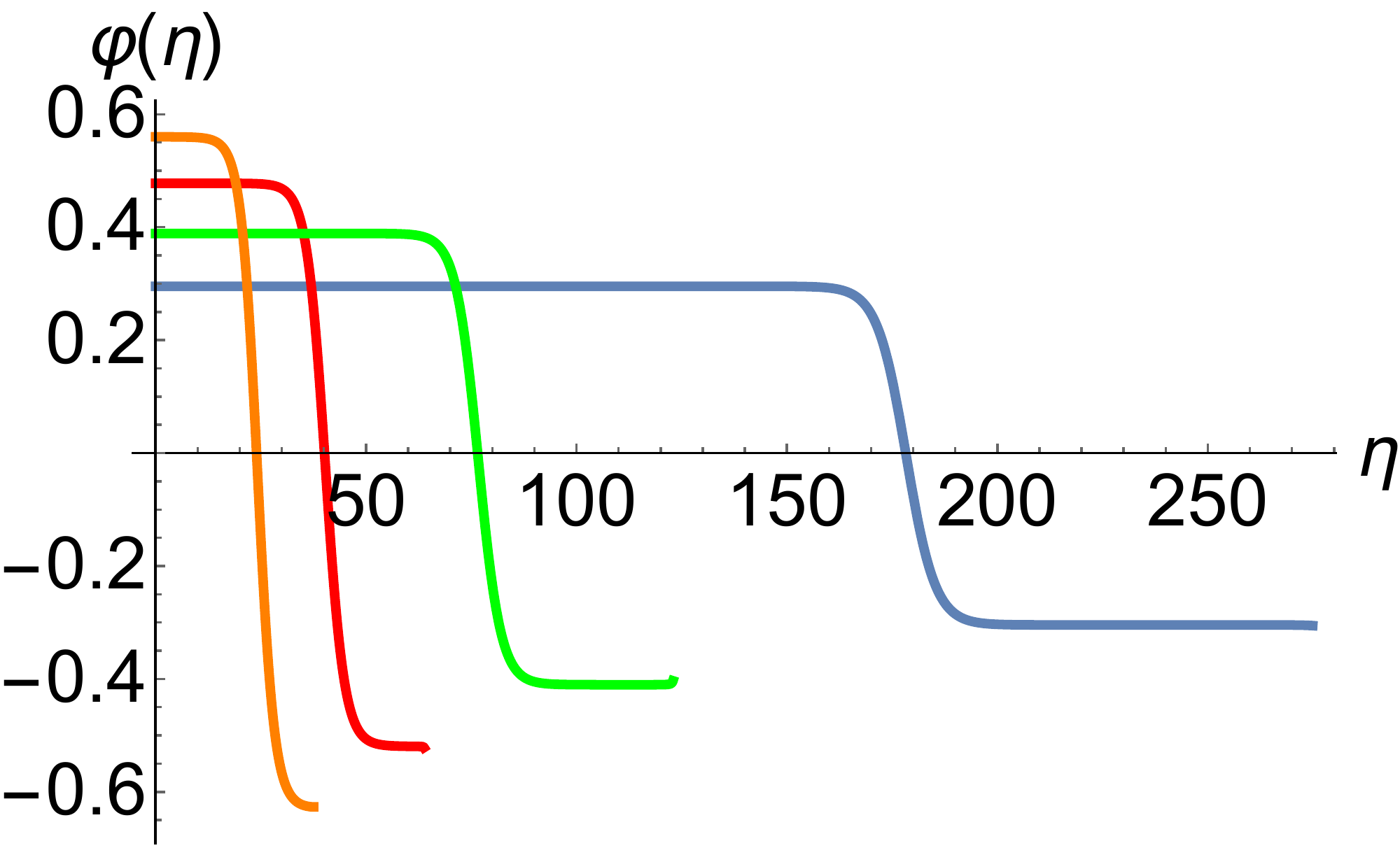}
	\caption{Plotted here is the evolution of four instantons in the potential given by equation (\ref{eq:quartic-potential}) 
but for four different values of $\mu$. The orange, red, green, and blue curves correspond to $\mu = 3/5, 1/2, 2/5,$ and $3/10$ respectively. 
\textit{Left:} The evolution of the scale factor in terms of Euclidean time $\eta$. \textit{Right:} The evolution of the scalar field.} \label{fig:analytic-comparison-cdl}
\end{figure}

\begin{figure}[ht!]
	\centering
	\includegraphics[width=0.6\textwidth]{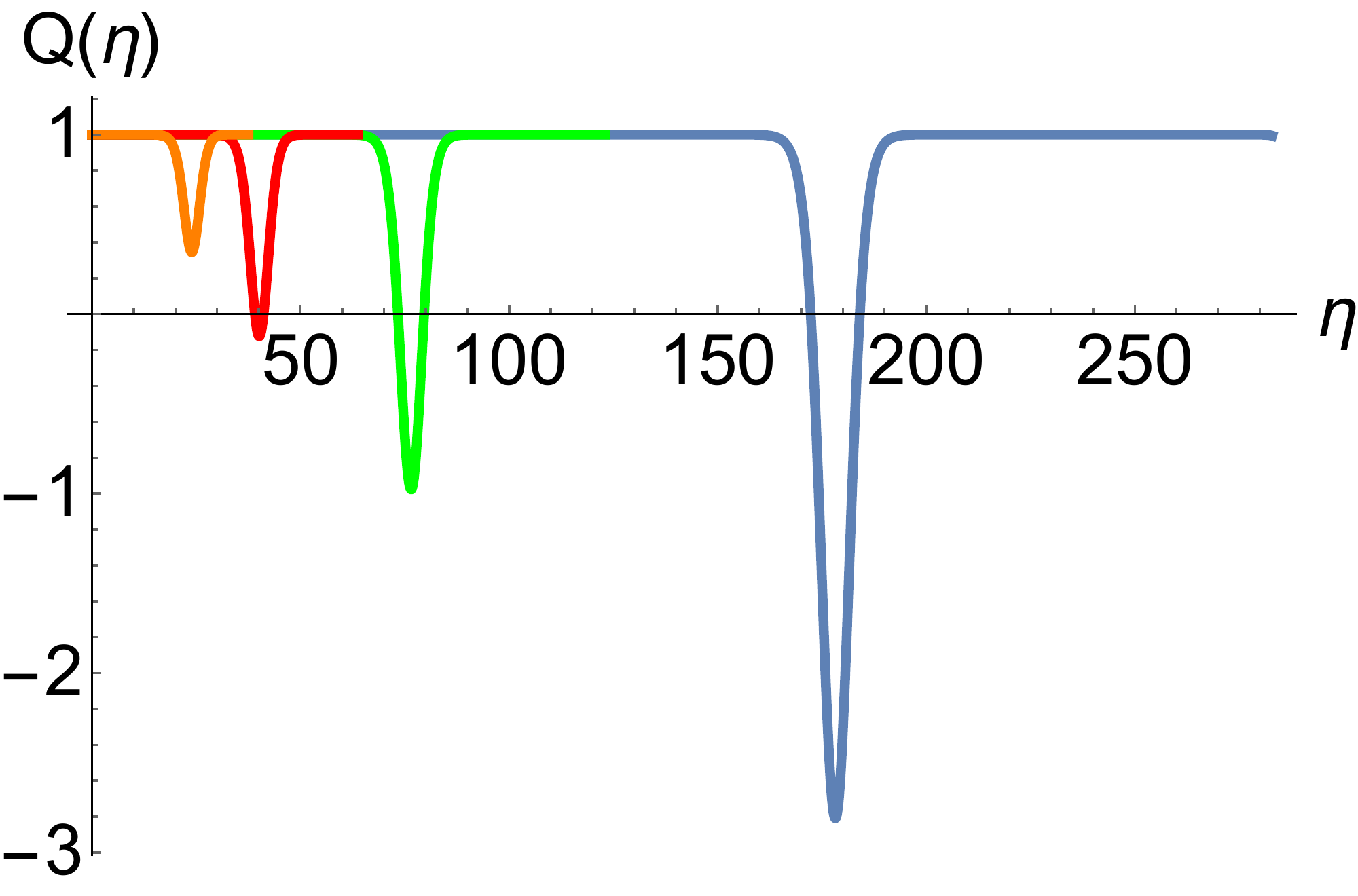}
	\includegraphics[width=0.35\textwidth]{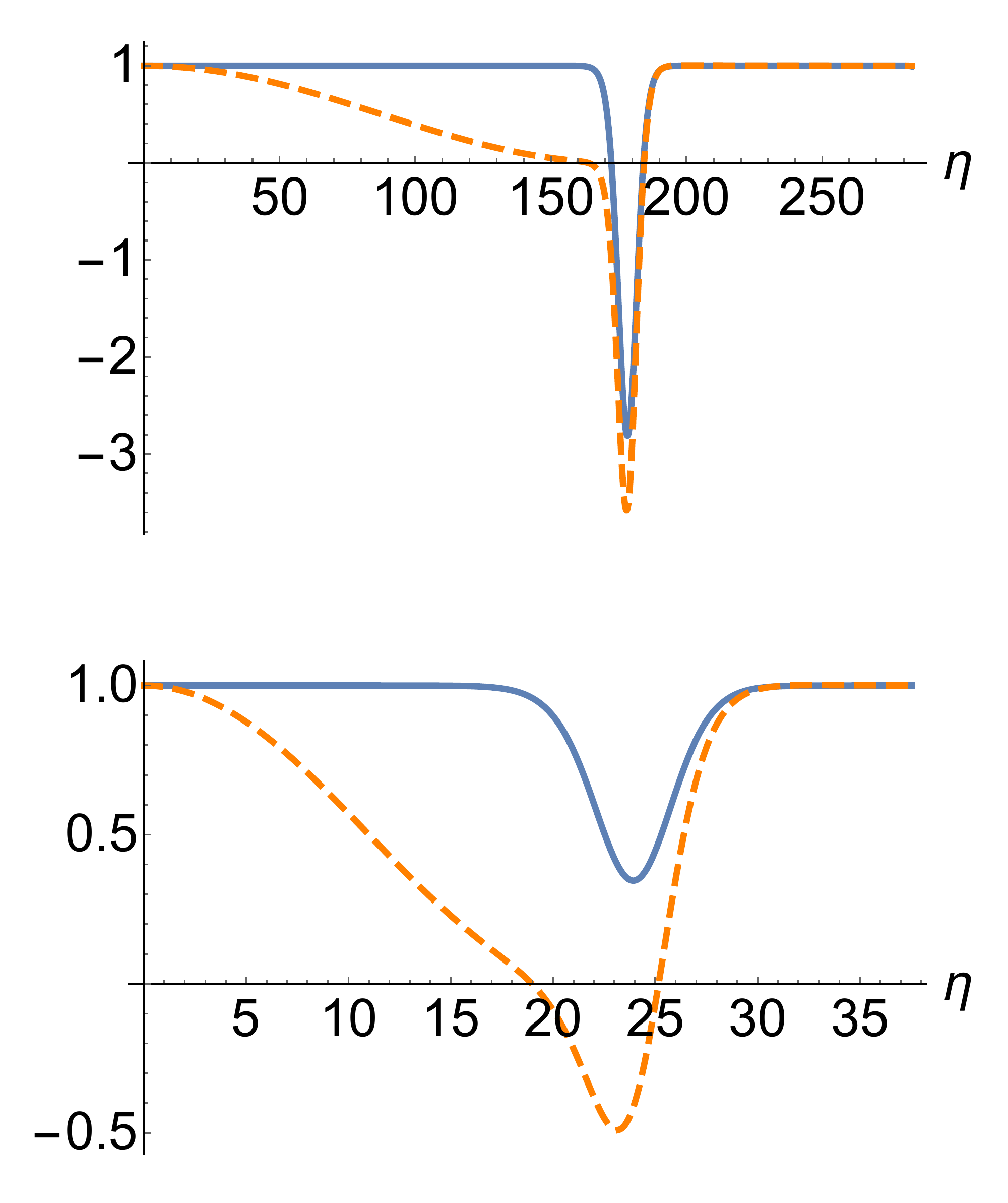}
	\caption{\textit{Left:} The kinetic pre-factor $Q$ for the bounces shown above. \textit{Right:} Comparison of $Q$ in blue and $Q_{L}$ in dashed orange. At the top $\mu = 3/10$ while at the bottom $\mu = 3/5$.} \label{fig:analytic-comparison-q}
\end{figure}

In deriving the analytic bounds for $\mu$ we took several approximations. Therefore it is useful to compare the approximate analytics to the full, numerical solutions. Here we choose $\kappa = c = 1$ for simplicity and without loss of generality and compare the two methods for various values of $\mu$. Note that since $\epsilon$ scales like $\mu^6$, the thin wall approximation is satisfied very rapidly as $\mu$ decreases from 1. Four sample geometries are shown in Fig.~\ref{fig:analytic-comparison-cdl} while their corresponding $Q$ values are plotted in Fig.~\ref{fig:analytic-comparison-q}. In table \ref{tab:analytics-vs-numerics} we compare the analytics with the numerics, indicating that our approximation yields excellent results. In particular, the approximation of taking $\varphi_c = 0$ is a very good one while the largest uncertainty comes from neglecting the derivative of $\rho$. From Fig. (\ref{fig:analytic-comparison-q}) is also apparent that the Hamiltonian kinetic pre-factor $Q$ and its Lagrangian counterpart $Q_{L}$ behave in a very similar fashion when $\mu$ is large but may differ qualitatively in other situations. In particular since $Q_{L}$ always develops a negative region, the difference between the two grows as $\mu$ shrinks.

\begin{table}[ht!]
	\centering
	\begin{tabular}{lll|ll|ll|ll}
		& \multicolumn{2}{c}{$\mu = 3/5$} & \multicolumn{2}{c}{$\mu = 1/2$} & \multicolumn{2}{c}{$\mu = 2/5$} & \multicolumn{2}{c}{$\mu = 3/10$}  \\
	    & Numerics & Analytics & Numerics & Analytics & Numerics & Analytics & Numerics & Analytics  \\ \cline{1-9}
		$\rho'_c$  & -0.4901 & 0 & -0.4939 & 0 & -0.4982 & 0 & -0.4976 & 0\\
		$\phi_c$ & 0.0108 & 0 & 0.0037 & 0 &  -0.0001 & 0 & 0.0002 & 0  \\
		$\rho_c$ & 13.266 & 14.001 & 23.250 & 24.132 &  45.927 & 47.036 &  109.852 & 111.323 \\ \cline{1-9}
		$\rho_m$ & 13.898 & 14.001 & 24.019 & 24.132 & 46.916 & 47.036 &  111.199 & 111.323 \\ \cline{1-9}
		$Q_{min}$ & 0.3457 & $\leq$ 0.4583 & -0.1242 & $\leq$ 0 & -0.9768 & $\leq$ -0.8437 &  -2.8087 & $\leq$ -2.6667
	\end{tabular} \label{tab:analytics-vs-numerics}
	\caption{Comparison of various quantities in the analytic expression with the numerics. The ones with subscript c refer to the the values where $Q$ takes the minimum. $\rho_m$ is the maximum/critical bubble radius and $Q_{min}$ is the minimum value for $Q$.}
\end{table}
These results are still of order one in $\mu$ which corresponds to a field excursion for $\phi$ of order one also which might be considered problematic. On the other hand, the approximations we are using work better for ever smaller values $\mu$, hence even though it is numerically very hard to find Coleman - De Luccia instantons for these values, we can nevertheless rely on the analytical tools developed to analyze these solutions.

\section{Negative mode problem for Higgs-like potentials} \label{Higgs}
\begin{figure}[ht!]
	\centering
	\includegraphics[width=0.6\textwidth]{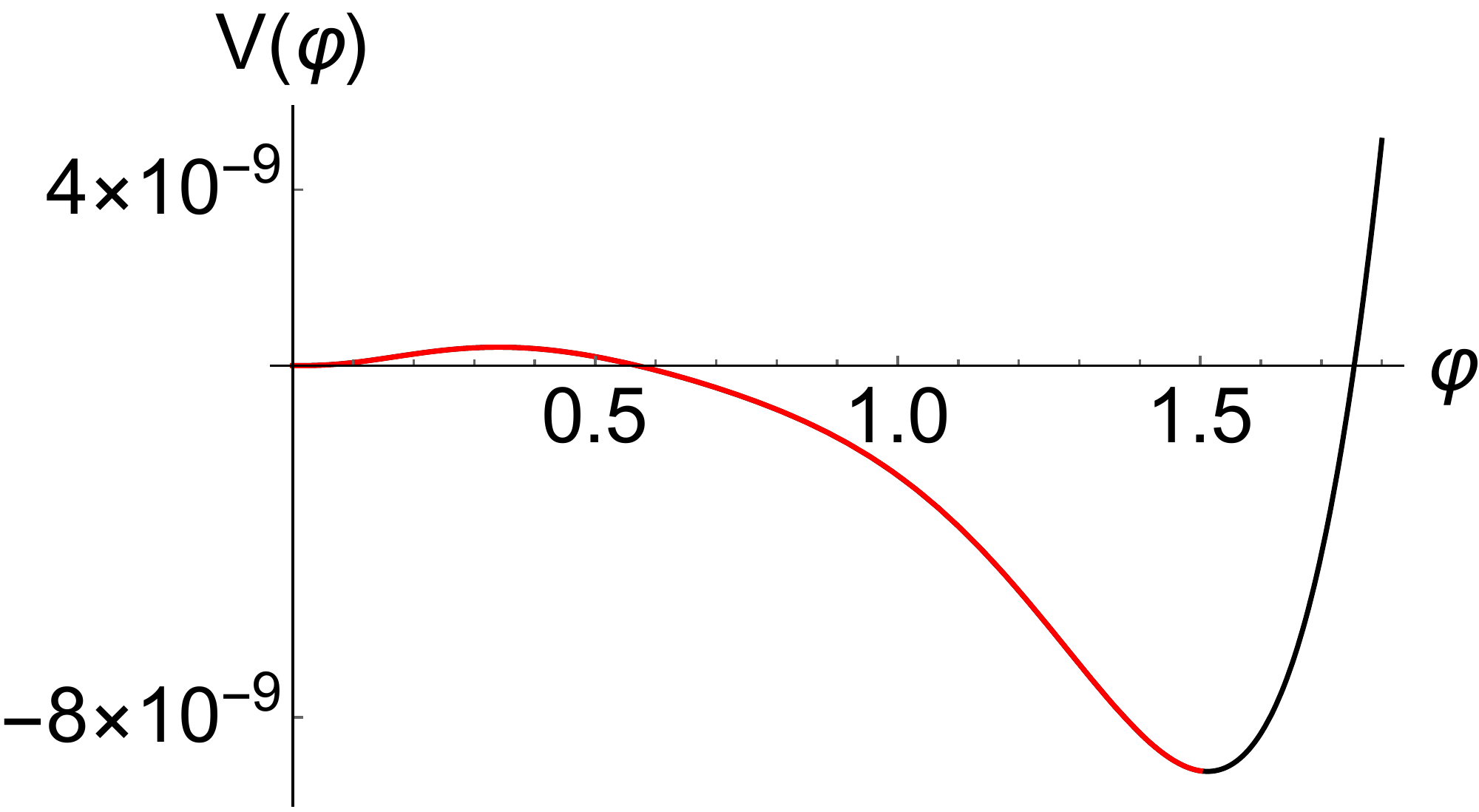}
	\caption{An example of the Higgs-like potential described in Eq. (\ref{eq:higgs-potential}) for $q = 10^{-7}$ and $\Lambda = 0.57$. The bounce solution is marked in red and does not develop a problematic, negative $Q$, region.} \label{fig:higgs-pot}
\end{figure}

\begin{figure}[ht!]
	\includegraphics[width=1.1\textwidth]{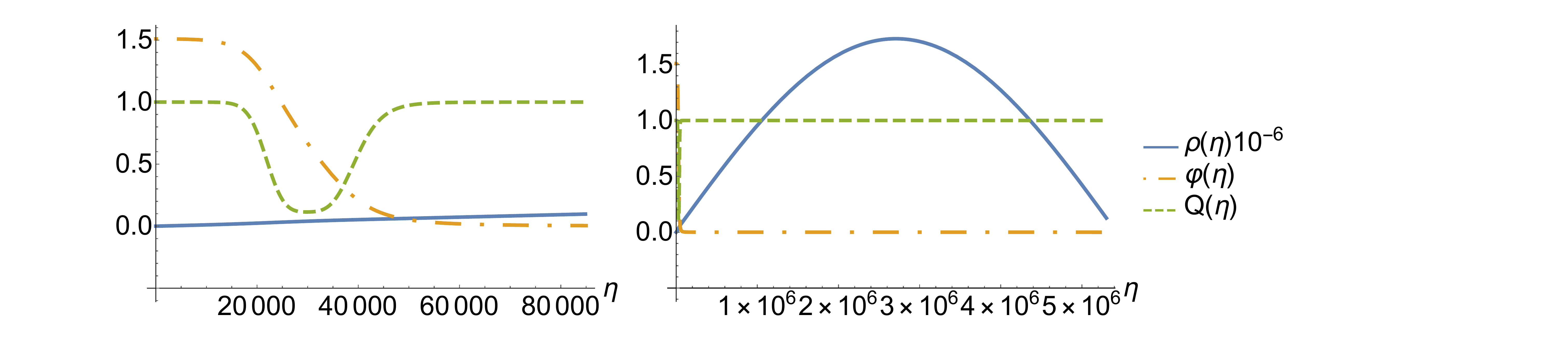}\\
    \includegraphics[width=1.1\textwidth]{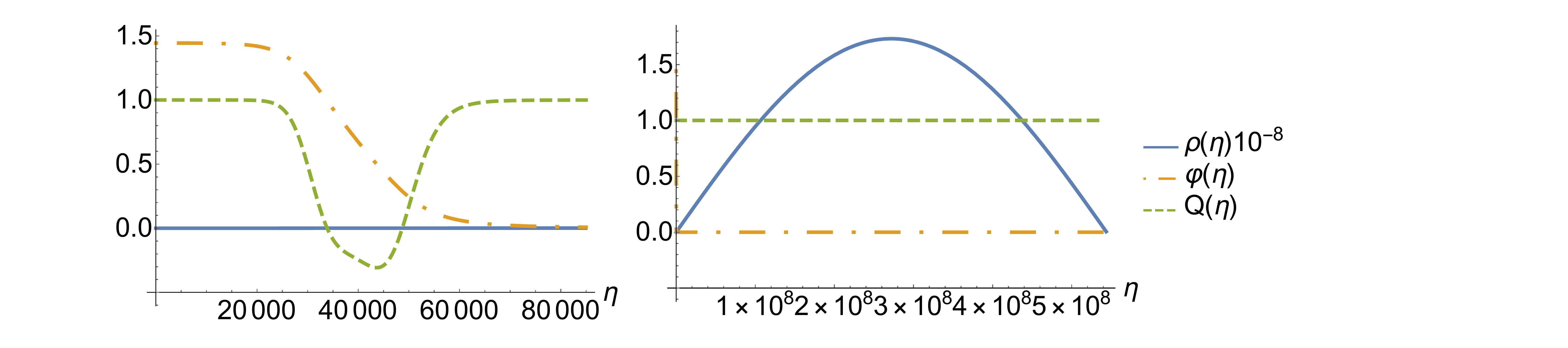}
	\caption{Here we show the values of the scalar field $\varphi$, scale factor $\rho$ and the function $Q$ for
the Higgs like potential Eq.~(\ref{eq:higgs-potential}).
The top figure shows the Coleman - De Luccia 
instanton for $\Lambda = 0.57$  while the bottom one has $\Lambda = 0.6$.
The images on the left are zoomed in versions of the full instantons shown on the right.
$M_{Pl} = 1$ units are used where we zoomed in on the part of the instanton where the scalar field tunnels and the problematic behaviour of $Q$ might occur.} \label{fig:higgs}
\end{figure}

Taking into account the current experimental bounds of the standard model parameters, the instability scale of the Higgs potential, $\lambda(\mu_{\Lambda})=0$,
depends sensitively on the top Quark and Higgs masses. The bounds at 1$\sigma$ currently are \cite{Markkanen:2018pdo}
\begin{align}
1.16 \cdot 10^{9}~\text{GeV} < \mu_{\Lambda} < 2.37 \cdot 10^{11}~\text{GeV} \pkt
\end{align}
such that the top of the potential barrier lies at about
\begin{align}
\varphi_{top} = 4.64 \cdot 10^{10}~\text{GeV} \kma
\end{align}
and the barrier height is
\begin{align}
V_{top} = 3.46 \cdot 10^{38}~\text{GeV}^{4} = (4.31 \cdot 10^{9}~\text{GeV})^4 \pkt
\end{align}
In Planck units $M_{Pl}= 1/\sqrt{8\pi G} \approx 2.435 \cdot 10^{18}~\text{GeV} = 1$, these numbers are:
\be \label{eq:instability scale}
4.76 \cdot 10^{-10} < \mu_{\Lambda} < 9.73 \cdot 10^{-8} \kma
\varphi_{top} = 1.91 \cdot 10^{-8} \kma
V_{top} = 9.84 \cdot 10^{-36} \pkt
\ee
At high energies the Higgs potential can be modelled as \cite{Gregory:2018bdt}
\begin{align} \label{eq:higgs-potential}
V_{H}= V_0 + \frac{\lambda_{H}(\varphi)}{4} \varphi^4  \kma  \\
\lambda_{H}= q \left[({\rm ln}\varphi)^4-({\rm ln}\Lambda)^4 \right] \kma
\end{align}
where $q$ is a dimension-less fitting parameter and $V_0$ is the cosmological constant. An sample potential for specific values of $q$ and $\Lambda$ is given in Fig. (\ref{fig:higgs-pot}). We can further mimic the Higgs potential by choosing $V_0 << V_{top}$ and
\begin{enumerate}
	\item $\Lambda = 10^{-9}, q=10^{-2}$ for the lower bound value of instability scale or
	\item $\Lambda = 10^{-7}, q=10^{-9}$ for the upper bound value of the instability scale, Eq.~(\ref{eq:instability scale}).
\end{enumerate}
Numerically, we found that for $\Lambda < \Lambda_*$ $Q$ is positive everywhere while
for $\Lambda > \Lambda_*$, $Q$ develops a region with $Q<0$.
Choosing parameters $q = 10^{-7}$ and $V_0 = 10^{-12}$ we found $0.57 < \Lambda_* < 0.6$, see Figure \ref{fig:higgs}. Therefore for a realistic Higgs like potential, the negative mode problem shows up only at the Planck values of the instability scale.

\section{Concluding remarks} \label{conclusion}
Using the Hamiltonian approach to false vacuum decay \cite{Khvedelidze:2000cp, Gratton:2000fj},
we have shown that for generic polynomial potentials the negative mode problem is not related to Planck scale physics. At the same time we demonstrated that for a Higgs - like potential, a region with $Q<0$ does not develop for realistic values of the potential's parameters. Instead, the problem only shows up if we assume the Higgs instability scale to be close to the Planck mass.

In the present analysis we used the Hamiltonian reduction scheme, which is based on Dirac's approach to constrained dynamical systems. Within this method, both, gauge fixed \cite{Khvedelidze:2000cp} and gauge invariant \cite{Gratton:2000fj} approaches, are not problematic and give the same answer. Hence we think this reduction gives a more adequate description of the physical situation than the Lagrangian approach. Note that there is a similar controversy in the counting of the number of negative modes \cite{Rubakov:1996cn}, \cite{Alonso:2017avz} of axionic Euclidean wormholes \cite{Lavrelashvili:1987jg,Giddings:1987cg}. Recently it was advocated that the Hamiltonian approach discussed here, also gives the correct answer in the wormhole case \cite{Hertog:2018kbz}.
On the other hand why Lagrangian and Hamiltonian reductions give a different kinetic pre-factor $Q$ for bounces in false vacuum decay and its physical relevance is still an open, puzzling question. It will be exciting to see if the implementation of a more general framework by not only considering Euclidean but a fully complex lapse as was proposed in \cite{Bramberger:2016yog} and applied in a cosmological setting in \cite{Bramberger:2017cgf} could resolve this issue. Another interesting issue is to investigate in which realistic cosmological or astrophysical set up a situation with negative $Q$ could occur and what the physical consequences might be. We hope to return to these questions in further study.

\acknowledgments

We thank Jean-Luc Lehners for stimulating discussions.
The work of SFB is supported in part by a fellowship from the Studienstiftung des Deutschen Volkes.
The work of G.L. is supported in part by the Shota Rustaveli National Science Foundation of Georgia with travel Grant MG-TG-19-117.

\bibliographystyle{utphys}
\bibliography{bib-negativeQ}

\end{document}